\documentstyle[twoside,fleqn,espcrc2]{article} \input{epsf}

\font\zz=cmss10

\newcommand{\plaquette}{
\mathop{{\vcenter{\hrule height .5pt
\hbox{\vrule width .5pt height 4pt \kern 4pt
    \vrule width .5pt} \hrule height .5pt}}}}

\newcommand{\Z}{\hbox{\zz Z} \kern-.4em \hbox{\zz Z}}

\title{Gauge fixing and Gribov copies in pure Yang-Mills on a circle.}

\author{James E. Hetrick\address{Institute for Theoretical Physics,
        University of Amsterdam \\ Valckenierstraat 65, 1018 XE
Amsterdam, The Netherlands}}

\begin{document}

\begin{abstract}
In order to understand how gauge fixing can be affected on the
lattice, we first study a simple model of pure Yang-mills theory on a
cylindrical spacetime [$SU(N)$ on $S^1 \times$ {\bf R}] where the
gauge fixed space is explicitly displayed. On the way, we find that
different gauge fixing procedures lead to different Hamiltonians and
spectra, which however coincide under a shift of states.  The lattice
version of the model is then compared.
\end{abstract}
\maketitle

\section{Introduction}

A crucial question for understanding lattice gauge fixing, is how the
reduction of a gauge theory \cite{GitTy} is
carried over from the continuum to the discretized version by the
Wilson formulation. Given a theoretical understanding of this
processs, efficient algoritms can be explored.

The analysis here continues a series of steps \cite{GFL,SonofGFL} in
understanding the topological structures involved and their remains
upon discretization.  Here we present the gauge fixing
of a simple model where the above steps can be explicitly
displayed analytically.

For $SU(N)$ gauge fields in $1+1$ dimensions where space is
compactified to a circle \cite{Rajeev,HH,Mklson,Witten,WittenRev,LS},
the spectrum is discrete and depends on the choice of gauge fixing.

In general, two dimensional Yang-Mills theory on a Riemann surface is
a topological theory, whose analysis is presently the subject of much
research \cite{Witten,Danny}. These theories have only
non-perturbative excitations which depend on the underlying topology
of their spacetime. In two dimensions, Yang-Mills field theory becomes
a finite dimensional quantum mechanics of the eigenvalues of the
holonomies (Wilson loops) around each non-contractible circle.
The addition of fermions has an interesting effect on the vacuum
structure as pointed out in
\cite{HH,LS,LS2}, however we will confine ourselves to the pure gauge
case in this talk for the purpose of displaying the gauge fixing
constructs.

The fundamental fields of the theory are the gauge fields $A_\mu(x,t)$
defining the coordinates of the unreduced configuration space, with
the usual Lagrangian,
\begin{equation}
{\cal L} = -{1\over 2}{\rm Tr}~F_{\mu\nu}F^{\mu\nu},
\end{equation} 
The fields $A_\mu$ and the curvature $F_{\mu\nu}$ take their values in
the Lie algebra $\widehat g$ of $G = SU(N)$.  On the circle we can,
without loss of generality, take $A_\mu(x+L,t)= A_\mu(x,t)$.

The canonical momenta to the $A_\mu$ are:
\begin{equation}
\Pi_\mu^a = E_\mu^a = {\delta {\cal L}\over \delta \partial_0 A_\mu^a},
\end{equation} 
and we see immediately that $\Pi_0$ vanishes identically.

The classical Hamiltonian is then
\begin{equation}
H = {1\over 2}\int_0^L dx \{E^2_1 + 2A_0 D_1 E_1\}
\end{equation} 
and $A_0$ emerges as a Lagrange multiplier field for the constraint of
Gauss' law
\begin{equation}
D_1 E_1 \sim 0
\end{equation} 
which must vanish on the reduced phase space ($\sim 0$).  Thus we can
set $A_0 = 0$ in further discussion.

As a secondary constraint, $D_1 E_1$ generates gauge
transformations on the partially reduced phase space \cite{GitTy}
which we must reduce further to the equivalence classes under gauge
transformations.

\section{Gauge Fixing}

The configuration space of the theory is now ${\cal A} =
\{A_1(x)~|~A_1: S^1 \mapsto \widehat g\}$, the set of maps of the
circle into $\widehat g$. Similarly the space of gauge transformations
is the set of maps from the circle into the gauge group, ${\cal G} =
\{\Lambda(x)~|~\Lambda: S^1 \mapsto G\}$. The true configuration phase
space is then formally
\begin{equation}
{\cal A}/{\cal G}.
\end{equation} 

What makes this circle model so convenient is the fact that ${\cal A}$
and ${\cal G}$ have been well studied
\cite{Loops}. ${\cal G}$ is the loop group of $G~(\equiv LG)$, whose
definition is given above, and $\cal A$ is isomorphic to the tangent
space of $LG~(\equiv L\widehat g)$.

$LG$ has the following structure:
\begin{equation}
LG(={\cal G})/G ~~\equiv \Omega G
\end{equation} 
where $\Omega G$ is the set of {\it based} loops in $G$, that is, the
set of maps $\{ \Lambda: S^1 \mapsto G \}$ such that $x=0 ~(\in S^1)
\to ${\bf 1} $(\in G)$.

It is known that \cite{Mklson,Loops,AtiyahBott}
\begin{equation}
L\widehat g (={\cal A})/\Omega G \cong G.
\end{equation} 
Then, using eqs. (6) and (7) in (5) gives:
\begin{equation}
{\cal A}/{\cal G} \cong G/G.
\end{equation}  
where $G/G$ is obtained by identifying the points $x^\prime$ and $x
\in G$ which are conjugate: $x^\prime = a^\dagger x a,~\forall a$,
under the action of $G$ on itself. This space is an orbifold made by
identifying points in the maximal Abelian subgroup of $G$ (the maximal
torus $T_G$), under the action of the Weyl group $W_G$, a discrete set
of transformations which shifts an element of $T_G$ by the center
$Z_N$, or permutes its diagonal elements. Thus
\begin{equation}
{\cal A}/{\cal G} \cong T_G/(Z_N \times S_N),
\end{equation}  
and we have an exact identification of the physical configuration
space of the theory as the orbifold $T_{SU(N)}/(Z_N
\times S_N)$.

\subsection{Coulomb Gauge}

The above analysis gives the topology of the gauge orbit space.
How do we see this structure emerge in the more colloquial method of
gauge fixing, say to Landau or Coulomb gauge conditions?

In the Coulomb gauge \cite{HH,LS} for example, we must find a gauge
transformation $\Lambda(x,t)$ such that
\begin{equation}
\partial_1\{~\Lambda A_1\Lambda^\dagger
+ ig\Lambda\partial_1\Lambda^\dagger \} = 0,~~~~\forall A_1(x,t).
\end{equation}  
Eq. (10) is solved by taking
\begin{equation}
\Lambda = e^{-iBx/L}~\exp\{\int_0^xdyA_1(y,t)\}
\end{equation}  
where
\begin{equation}
B = B(t) = \int_0^LdyA_1(y,t)
\end{equation}  
is the zero momentum mode of $A_1$. $\exp\{-iB(t)\}$ is the
holonomy around the circle. Furthermore, within the Coulomb gauge
condition we have the freedom to make $x$-independent gauge
transformations, and so we can diagonalize $A_1(t)$ on each time slice
of the cylinder:
\begin{equation}
A_1(t) \rightarrow {1\over gL}\left(
\begin{array}{c} \theta_1(t)~~~~~~~~  \\
                     \ddots \\ ~~~~~~~~~~~ \theta_N(t)
\end{array}\right) \equiv
\Theta(t).
\end{equation}  

As $A_1$ is traceless, $\sum_{i=1}^N\theta_i(t) = 0$.
In this complete Coulomb gauge one can easily show that $A_0$
decouples \cite{HH}.

Now we see the physical configuration space emerging, since $\Theta$
is an $N-1$ dimensional flat sub-manifold of ${\cal A}$, which will
become the maximal torus $T_{SU(N)}$ upon the identification Gribov
copies.

\subsection{Gribov Copies}

Notice that maintaining the periodicity of $A_\mu(x)$ around the
circle, requires $\Lambda$ to satisfy:
\begin{equation}
\Lambda(x+L,t)\Lambda^\dagger(x,t) = Z_N
\end{equation}  
where $Z_N$ is in the center of $SU(N)$. Thus eq. (13) is preserved
under gauge transformations
\begin{equation}
\Lambda(x)_{jk} = \exp\{{2\pi ix\over L}(n_j + m)\}\delta_{jk}
\end{equation}  
where $n_j$ and $m$ satisfy
\begin{eqnarray}
&n_j\in \Z,~~~~~\sum_{j=1}^N n_j = 0 \\ 
&m=\kappa/N,~~~~~\kappa = 1,\dots,N-1
\end{eqnarray}  
such that the action on $\Theta$ is
\begin{equation}
\theta_j(t) \rightarrow \theta_j(t) + 2\pi n_j + 2\pi m.
\end{equation}  

The identification of the points $n_j$ in the space spanned by
$\Theta$ make it isomorphic to the maximal torus $T_{SU(N)}$, while
the Gribov copies labeled by $m$ make the $Z_N$ identifications.
Similarly, a further set of Gribov copies $\Lambda(t)$ permutes the
$\theta_j(t)$.  These then, are all the residual Gribov copies left
after completely fixing to Coulomb gauge. Thus the identification of
these copies gives the linear subspace $\Theta$, coming from Coulomb
gauge fixing, the topology of the physical configuration space,
$T_{SU(N)}/(Z_N\times S_N)$.

The Gribov copies are essential to identifying the correct topology of
the configuration space which manifests, for instance, in the
periodicity of the wave function $\Psi(\theta_j)$.  This argument has
also been stressed in \cite{Pierre} and \cite{LS2}.

\section{Quantization}

The Hamiltonian, on the unreduced configuration space, must be
projected down onto the gauge fixed surface (via Dirac brackets or
geometric methods). In the case at hand,
\begin{equation}
H = \int_0^Ldx~\Pi_1^2 = \int_0^Ldx~{\delta^2 \over \delta A_1^2(x)}
\end{equation}  
and there are at least two projections in the literature.

The method in \cite{HH} and \cite{LS} is to take the
restriction of eq.  (19) to the hyperplane in ${\cal A}$ spanned by
$\Theta$, then impose the periodicity of the Gribov copies on the wave
funtions. For $SU(2)$,
\begin{eqnarray}
H = {g^2L\over 8}{\partial^2\over \partial
\theta^2}~~~~~~~~~~~~~~~~~~~~
\\  
\Psi(\theta) = \cos(n \theta), ~~E_n = g^2Ln^2/8; ~~n\in\Z_{\geq 0}
\end{eqnarray}  

In \cite{Rajeev} and \cite{Mklson}, the full Hamiltonian in eq. (19)
is projected first onto the subspace of ${\cal A}$ isomorphic to $G$,
identified there with the Laplacian on $G$, and the equivalence of
conjugate $(G/G)$ configurations is imposed on the wave functions.
Since $H = \Delta_{SU(N)}$, wave functions are the characters of
$SU(N)$ and the spectrum is the quadratic Casimir.  Thus for $SU(2)$
again:
\begin{eqnarray}
\Psi(\theta) = \sin([j+{1\over 2}]\theta)/\sin(\theta/2)\\
E_n = g^2Lj(j+1)/2~~~~~j\in \Z_{\geq 0}/2.
\end{eqnarray}  

It would seem that these spectra are completely different, however
notice that adding ${g^2L\over 8}$ to the Casimir spectrum reproduces
the spectrum of eq. 8, up to the disappearance of the ground state
with $E_{n=0} = 0$ since
\begin{equation}
j(j+1)/2 + 1/8 = n^2/8;~~~~\forall n\in\Z_{>0},~j\in {\Z_{\geq 0}\over 2}
\end{equation}   
Similarly, for $SU(3)$, the quantization of a particle on the maximal
torus $T_{SU(3)}$ gives \cite{HH,LS}:
\begin{equation}
E_{\lambda,\mu} = {g^2L\over 6}(\lambda^2 + \lambda \mu + \mu^2);~~~~
\lambda,\mu\in\Z
\end{equation}
Taking $\lambda\rightarrow \lambda+1$, $\mu\rightarrow \mu+1$ yields
$c_2/2 + 1/2$, where the quadratic Casimir of $SU(3)$ is $c_2 =
(\lambda^2 + \lambda \mu + \mu^2)/3 + \lambda + \mu$.

The addition of certain constants (and higher Casimirs) to the
spectrum has also been noted by Witten \cite{WittenRev} and appears to
be related to the relationship of 2-D Yang-Mills to the equivalent
topological $BF$ theory. Here the added constant is one half the
scalar curvature of $G$, coming from the relation between the
Hamiltonians on $G$ and $T_{SU(N)}$ \cite{Dowker}.

The disappearance of the ground state wave function $\Psi_{n=0}$ in
the identification of the spectra is due to the identification of
wave functions from $G$ to $T_{SU(N)}$, which involves the metric of
$G$ \cite{InPrep}. This becomes non-trivial when fermions are included
\cite{HH,LS2}.

\section{Lattice Version}

In two dimensions the lattice partition function $Z$ of pure
Yang-Mills theory factorizes into ``pant diagrams'' which can be glued
together by integrating over the holonomy $W$ at the common boundary
\cite{Witten,Rusakov,Wheater,BT} as shown in figure 1.
\begin{figure}[htb]
\epsfxsize=65truemm
\centerline{\epsffile[147.4 496.1 467.7 694.5]{pants.ps}}
\caption{The partition function with external states $W_a,
W_b, W_c,$ and $W_d$:~ $Z_{1+2} = \int dW
{}~Z_1(W_a,W_b,W^\dagger)Z_2(W,W_c,W_d)$}
\label{fig:1}
\end{figure}
After integrating out all internal links, the contribution of the pant
diagram $Z_1(A_1;W_a,W_b,W)$ to $Z$ is
\begin{equation}
Z_1 = \sum_r d_r^{-1}\lambda_r^{A_1/a^2}\chi_r(W_a)\chi_r(W_b)
\chi_r(W)
\end{equation} 
where $r$ runs over representations of $G$, of dimension $d_r$;
$\chi_r(W_i)$ is the character of the Wilson loop $W_i$ in
representation $r$, $A_1$ is the area of the Riemann surface $Z_1$,
and $a$ is the lattice spacing. $\lambda_r$ is the Fourier component of
the character expansion of the lattice action
\begin{equation}
\lambda_r = d_r^{-1}\int dU_{\plaquette}~\chi_r(U_{\plaquette}) \exp\{\beta_0N
{\rm Tr}~U_{\plaquette}\}.
\end{equation}  

With the $n$-point pant diagrams one can analytically calculate the
expectation value of a Wilson loop in the representation $R$, at the
seam $W$, by inserting an extra character $\chi_R(W)$ into the
computation of the partition function $Z_{1+2}$, in figure 1.

To see which of the above spectra the non-gauge-fixed lattice choses,
we compute
\begin{equation}
<W(t)W(0)> = \sum_n|<0|\widehat W|n>|^2 e^{-tE_n}
\end{equation}  
by gluing 3 cylinders in series, with Wilson loop characters
sandwiched in between, and setting $W_{\rm ends} = {\bf 1}$.  Taking
the areas of the external cylinders very large, one finds that
$<W(t)W(0)> ~\sim \exp(-c_2t/2)$ in accordance with the Casimir
spectrum of eq. (23). This is, of course, not surprising since the
lattice formulation is compact, hence the quantization is done on $G$
as in eq. (23). Similarly, one can use a Hamiltonian approach, with
$A_0 = 0$, finding again a Casimir spectrum, for the same reasons of
compactness.

It appears that the lattice (or quantization on the group manifold)
cannot see the true ground state of this model, that of a constant
wave function on the gauge orbit space $T_{SU(N)}/(Z_N\times S_N)$.
The lattice ground state is a constant $\Psi_{j=0}$ on $G$, which
projects to the $\Psi_{n=1}$ state on the maximal torus \cite{InPrep}.

\section{Acknowledgements}

I wish to thank many people for stimulating and insightful
disscussions on the above topics, in particular: Y. Hosotani, J. Smit,
Ph. de Forcrand, R. Dijkgraaf, P. van Baal, D. Birmingham, F.A. Bais,
and M. Blau.


\end{document}